\documentclass[3p,compress,numbers]{elsarticle}
\usepackage{epstopdf}
\usepackage{subfigure,graphicx,tabularx}
\usepackage{float}
\usepackage{amsmath, amsfonts, amssymb,mathrsfs}
\usepackage{amsthm}
\usepackage{epsf}
\usepackage{amsfonts}
\usepackage{stmaryrd}
\usepackage{amssymb}
\usepackage{leftidx}
\usepackage{color}
\usepackage{mathtools}
\usepackage{placeins}
\usepackage{booktabs}
\usepackage{enumitem}
\usepackage{caption}
\usepackage{multirow}
\usepackage{stackengine}
\usepackage[utf8]{inputenc}
\usepackage[english]{babel}
\usepackage{color}
\usepackage{pifont}
\usepackage{fontawesome}
\usepackage{hyperref}
\usepackage{hhline}

\DeclareMathOperator{\sgn}{sgn}
\newcolumntype{?}{!{\vrule width 1pt}}

\DeclareMathAlphabet{\mathsfit}{T1}{\sfdefault}{\mddefault}{\sldefault}
\SetMathAlphabet{\mathsfit}{bold}{T1}{\sfdefault}{\bfdefault}{\sldefault}

\theoremstyle{plain}
\newtheorem{theorem}{Theorem}

\newtheorem{remark}[theorem]{Remark} 
\newtheorem*{definition}{Definition}

\def\bfu{{\bf u}}

\def\bfE{{\bf E}}

\def\bfI{{\bf I}}

\def\bfN{{\bf N}}

\def\bfX{{\bf X}}

\def\bfs{{\bf s}}

\def\e0{\varepsilon_0}

\def\s0{\sigma_0}

\def\sts{\sigma_{\texttt{ts}}}
\def\scs{\sigma_{\texttt{cs}}}
\def\shs{\sigma_{\texttt{hs}}}

\begin{document}
\begin{frontmatter}

\title{A Guide to Fully Characterize the Fracture Properties of Cementitious Materials from Simple Experiments\vspace{0.1cm}}

\vspace{-0.1cm}

\author{Subhrangsu Saha}
\ead{saha15@illinois.edu}

\author{Bruce J. Moore}
\ead{brucem2@illinois.edu}

\author{Ben Manaugh}
\ead{manaugh3@illinois.edu}

\author{Jeffery R. Roesler}
\ead{jroesler@illinois.edu}

\author{Oscar Lopez-Pamies}
\ead{pamies@illinois.edu}

\address{Department of Civil and Environmental Engineering, University of Illinois, Urbana--Champaign, IL 61801, USA}

\begin{abstract}

\vspace{-0.1cm}

Guided by recent advances in the understanding of nucleation and propagation of fracture in elastic brittle materials, this paper proposes a suite of three simple experiments that permit the measurement of the three macroscopic material properties governing when and where cracks nucleate and propagate in structures made of cementitious materials that are subjected to arbitrary monotonic quasi-static loading conditions. The first experiment is that of the uniaxial compression of a cylindrical specimen, which enables the extraction of the elastic properties --- namely, the Young's modulus and Poisson's ratio --- as well as the uniaxial compressive strength. The second experiment is the Brazilian fracture test, performed with flat platens on a material disk to determine the uniaxial tensile strength. Having knowledge of the uniaxial compressive and uniaxial tensile strengths then allows for the estimation of the strength surface of the material via interpolation (e.g., a Drucker-Prager fit). Finally, the third experiment is the wedge split test on a notched cube, which yields the fracture toughness. We demonstrate by means of direct comparisons with four-point and three-point bending tests on both unnotched and notched beams made of a 3D-printable mortar mixture that the elasticity, strength, and toughness properties obtained from the proposed tests are sufficient to predict the nucleation and propagation of fracture for any structure (granted separation of length scales) made of cementitious materials under any monotonic quasi-static loading condition.

\keyword{Fracture nucleation; Strength; Phase-field regularization; Mortar; Concrete}
\endkeyword

\end{abstract}

\end{frontmatter}

\section{Introduction}

In this paper, guided by recent advances in the fracture of nominally elastic brittle materials\footnote{Recall that an elastic brittle material --- the most basic type of idealization of a solid capable of fracture --- is a material that, in response to mechanical forces, either deforms elastically or creates surface, the latter being the sole mechanism by which it dissipates energy.} \cite{KFLP18,KRLP18,KLP20,KBFLP20,KLP21,KRLP22,KLDLP24,LDLP24,KKLP24,SDLP24,KLP25,KZDLP26,SRLP26,LPK25}, we propose three simple tests to measure the three macroscopic material properties that govern when and where cracks nucleate and propagate in cementitious structures subjected to arbitrary monotonic quasi-static mechanical loads. In doing so, we clarify a long-standing misunderstanding in the cementitious materials community by identifying the precise material properties that fully govern the phenomenon of fracture in mortar and concrete.

Before presenting the proposed tests in Section \ref{Sec: The experiments} and demonstrating in Section \ref{Sec: Demonstration} that the material properties that are extracted from them fully characterize the fracture properties of cementitious materials, we summarize below the three main advancements established in \cite{KFLP18,KRLP18,KLP20,KBFLP20,KLP21,KRLP22,KLDLP24,LDLP24,KKLP24,SDLP24,KLP25,KZDLP26,SRLP26,LPK25}, one at a time.

\subsection{A precise and complete definition of strength} 

The first of the three main results established in \cite{KFLP18,KRLP18,KLP20,KBFLP20,KLP21,KRLP22,KLDLP24,LDLP24,KKLP24,SDLP24,KLP25,KZDLP26,SRLP26,LPK25} is a precise and complete definition of strength, a macroscopic material property that has historically been subject to significant misinterpretation. As introduced in \cite{KLP20,KBFLP20}, the strength of an elastic brittle material is defined as follows.
\begin{definition}\label{Def-Strength} The strength of an elastic brittle material is the set of all critical stresses $\boldsymbol{\sigma}$ at which fracture nucleates in a specimen (of sufficiently large size to be regarded as a homogeneous continuum) when it is subjected to a state of monotonically increasing, spatially uniform, but otherwise arbitrary stress. Such a set of critical stresses defines a surface $\mathcal{F}(\boldsymbol{\sigma})=0$ in stress space --- potentially, any star-shaped surface containing the origin $\boldsymbol{\sigma}=\emph{\textbf{0}}$  --- which is referred to as the strength surface of the material.
\end{definition}

Note that this definition says nothing about when fracture nucleates in problems where the stress field is \emph{not} spatially uniform and so it says nothing about fracture nucleation at individual material points. This is fundamentally different from the still common --- though long experimentally debunked --- notions of strength found in the literature (since the pioneering work of Lam\`e \cite{Lame1831} in 1831), which incorrectly posit that violating a stress condition at a single material point triggers fracture nucleation. Indeed, over the past many decades, a multitude of experiments have repeatedly shown that fracture in a body where the stress field is \emph{not} spatially uniform does \emph{not} nucleate when the stress at a single material point violates a critical value. Instead, fracture nucleates once a stress violation has taken place over a sufficiently large region  of the body, the size and shape of this region being dependent on the geometry of the body, the material that is made of, and the specifics of the applied loading conditions.

The definition also says nothing about the strength surface $\mathcal{F}(\boldsymbol{\sigma})=0$ having to be of any particular form, other than it is star-shaped and contains $\boldsymbol{\sigma}=\textbf{0}$ in its interior. This is fundamentally different from the classical view which insists that fracture nucleation is governed by a universal stress criterion --- such as, for instance, maximum principal stress or maximum shear --- across all material classes; see, e.g., \cite{Lame1831,Rankine1858,Mohr1900}. This is not in agreement with experimental observations either. Instead, experiments show that different materials exhibit different strength surfaces. The reason for this diversity is clear. The strength surface $\mathcal{F}(\boldsymbol{\sigma})=0$ represents the macroscopic manifestation of the presence of microscopic defects, whose specifics are unique to different materials. Consequently, different types of defects result in different strength surfaces; see, e.g., \cite{BCLLP24,KZDLP26} for materials exhibiting vastly different strength surfaces.

For isotropic materials, such as mortar and concrete, the strength surface admits the isotropic representation
\begin{equation*}
\mathcal{F}(\boldsymbol{\sigma})=\widehat{\mathcal{F}}(\sigma_1,\sigma_2,\sigma_3)=0,
\end{equation*}
where $\widehat{\mathcal{F}}$ is a symmetric function of the principal stresses $\sigma_1$, $\sigma_2$, and $\sigma_3$. Without loss of generality, having demonstrated that it provides perhaps the simplest valid description of strength for a wide range of elastic brittle materials \cite{BCLLP24,KZDLP26}, we advocate in this paper for the use of a Drucker-Prager strength surface:
\begin{equation}\label{DP-1}
\mathcal{F}(\boldsymbol{\sigma})=\sqrt{J_2}+\dfrac{\sigma_{\texttt{cs}}-\sigma_{\texttt{ts}}}
{\sqrt{3}\left(\sigma_{\texttt{cs}}+\sigma_{\texttt{ts}}\right)} I_1-\dfrac{2\sigma_{\texttt{cs}}\sigma_{\texttt{ts}}}
{\sqrt{3}\left(\sigma_{\texttt{cs}}+\sigma_{\texttt{ts}}\right)}=0.
\end{equation}
Here, $I_1={\rm tr}\,\boldsymbol{\sigma}=\sigma_1+\sigma_2+\sigma_3$ and $J_2=\frac{1}{2}{\rm tr}\,\boldsymbol{\sigma}^2_{D}=\frac{1}{3}(\sigma_1^2+\sigma_2^2+\sigma_3^2-\sigma_1\sigma_2-\sigma_1\sigma_3-\sigma_2\sigma_3)$, with $\boldsymbol{\sigma}_{D}=\boldsymbol{\sigma}-\frac{1}{3}({\rm tr}\,\boldsymbol{\sigma})\bfI$, stand for two of the standard invariants of the stress tensor $\boldsymbol{\sigma}$, while the constants $\sigma_{\texttt{ts}}>0$ and $\sigma_{\texttt{cs}}>0$ denote the uniaxial tensile and compressive strengths of the material, that is, they denote the critical stress values at which fracture nucleates under uniform uniaxial tension $\boldsymbol{\sigma}={\rm diag}(\sigma>0,0,0)$ and uniaxial compression $\boldsymbol{\sigma}={\rm diag}(\sigma<0,0,0)$, respectively.

\begin{remark}\label{Remark-Strength} \emph{As demonstrated in \cite{KFLP18,KRLP18,KLP20,KBFLP20,KLP21,KRLP22,KLDLP24,LDLP24,KKLP24,SDLP24,KLP25,KZDLP26,SRLP26,LPK25} and as detailed below, the strength surface $\mathcal{F}(\boldsymbol{\sigma})=0$ plays a central role not only in the nucleation but also in the propagation of fracture in elastic brittle materials. Because this precise and complete definition of strength was not introduced until 2020, its previous absence (and, as discussed next, its interaction with the elasticity and fracture toughness) has been the primary bottleneck stalling progress in the understanding of fracture mechanics and, in turn, in the development of a predictive theory.}
\end{remark}

\subsection{The three macroscopic material properties governing fracture nucleation and propagation} 

The second main result established in the string of recent works \cite{KFLP18,KRLP18,KLP20,KBFLP20,KLP21,KRLP22,KLDLP24,LDLP24,KKLP24,SDLP24,KLP25,KZDLP26,SRLP26,LPK25} is that any potentially successful mathematical macroscopic description of crack nucleation and propagation in elastic brittle materials must account for three distinct macroscopic properties: ($i$) the elasticity of the material; ($ii$) its strength surface; and ($iii$) its fracture toughness or critical energy release rate. Specifically, for isotropic materials, a viable formulation must incorporate the isotropic elastic energy density
\begin{equation}\label{W-elastic} 
W(\bfE)=\dfrac{E}{1+\nu}\,{\rm tr}\,\bfE^2+\dfrac{E\nu}{(1+\nu)(1-2\nu)}({\rm tr}\,\bfE)^2
\end{equation}
along with the isotropic strength surface
\begin{equation}\label{F-strength}
\mathcal{F}(\boldsymbol{\sigma})=\widehat{\mathcal{F}}(\sigma_1,\sigma_2,\sigma_3)=0,
\end{equation}
and the scalar fracture toughness
\begin{equation}\label{Gc-toughness}
G_c.
\end{equation}
In these expressions, $\bfE(\bfu)=\frac{1}{2}(\nabla\bfu+\nabla\bfu^T)$ denotes the symmetrized gradient of the displacement field $\bfu$, or infinitesimal strain tensor, $E$ and $\nu$ are the Young's modulus and Poisson's ratio, and $\sigma_1$, $\sigma_2$, $\sigma_3$ are, once more, the principal stresses.

It is essential to recognize that the Young's modulus $E$, the Poisson's ratio $\nu$, the strength surface $\mathcal{F}(\boldsymbol{\sigma})=0$, and the fracture toughness $G_c$ are \emph{macroscopic material properties}. Accordingly, they are physically meaningful only when the body under consideration is sufficiently large relative to its underlying heterogeneities. In the case of mortar and concrete, in particular, these properties are valid only at scales significantly exceeding the maximum aggregate size, namely, 4.75 mm for mortar and 25 mm for concrete. While classical theoretical works have established that a length-scale separation factor of four is sufficient for the elasticity constants $E$ and $\nu$ to be treated as macroscopic material properties \cite{DW1996}, no such thresholds have yet been worked out for $\mathcal{F}(\boldsymbol{\sigma})=0$ and $G_c$. Experimental data on mass concrete \cite{Blanks1935} and mortar \cite{Gonnerman1925}, as well as our own 3D-printable mortar experiments indicate a stricter requirement: the smallest dimension of the body must be at least ten times larger than the largest heterogeneity to ensure the validity of the strength surface and fracture toughness as macroscopic material properties.

For nominally elastic materials, there is a range of direct and indirect experimental techniques that one can choose from to determine the elasticicty constants ($E$ and $\nu$), the strength surface ($\mathcal{F}(\boldsymbol{\sigma})=0$), and the fracture toughness ($G_c$). The selection of a specific method typically depends on the ease of specimen fabrication, the feasibility of applying the required loads, the practicality of measuring the resulting deformations, forces, and crack evolution, and, finally, the complexity of interpreting the data to extract the desired properties. 

As stated from the outset, the main objective of this work is to propose three simple tests to determine $E$, $\nu$, $\mathcal{F}(\boldsymbol{\sigma})=0$, and $G_c$ for cementitious materials. The proposed tests have been judiciously selected so as to provide a robust and effective methodology of broad utility to the cementitious materials community, with a particular emphasis on accessibility for practitioners. The tests and their corresponding analyses are presented in Section \ref{Sec: The experiments}.

\subsection{A complete regularized theory that explains, describes, and predicts nucleation and propagation of fracture} 

Building upon the two preceding results, the third main result established in \cite{KFLP18,KRLP18,KLP20,KBFLP20,KLP21,KRLP22,KLDLP24,LDLP24,KKLP24,SDLP24,KLP25,KZDLP26,SRLP26,LPK25} is a regularized phase-field formulation that provides a descriptive and predictive framework for crack nucleation and propagation in elastic brittle materials. Within this approach, a continuous order parameter, or phase field $v\in[0,1]$, is used to represent cracks, with $v=1$ indicating intact material and $v=0$ representing completely fractured zones. A regularization length scale $\varepsilon>0$ governs the smooth transition between these two states. Originally introduced in \cite{KFLP18}, this formulation was later cast into a thermodynamically consistent framework \cite{KRLP18}, proven mathematically well-posed \cite{LDLP24}, and extensively validated against experimental data \cite{KFLP18,KRLP18,KLP20,KBFLP20,KLP21,KRLP22,KLDLP24,KKLP24,KLP25,KZDLP26,LPK25}. 

The theory is expressed as a system of two coupled partial differential equations (PDEs) for the deformation field $\bfu(\bfX,t)$ and the phase field $v(\bfX,t)$, subject to appropriate initial and boundary conditions. When written in their time-discretized version, the displacement field $\bfu_k(\bfX)=\bfu(\bfX,t_k)$ and phase field $v_k(\bfX)=v(\bfX,t_k)$ at any material point $\bfX$ in the domain $\Omega$ occupied by the body of interest and at any given discrete time $t_k\in\{0=t_0,t_1,...,t_m,t_{m+1},...,t_M=T\}$ are determined by the system of equations
\begin{equation}
\left\{\begin{array}{ll}
\hspace{-0.15cm} {\rm Div}\left[v_{k}^2 \dfrac{\partial W}{\partial \bfE}\left(\bfE(\bfu_{k})\right)\right]+v_k^2 \textbf{b}(\bfX,t_k)=\textbf{0},& \bfX\in\Omega\vspace{0.2cm}\\
\hspace{-0.15cm} \bfu_k(\bfX)=\overline{\bfu}(\bfX,t_k), & \bfX\in\partial\Omega_{\mathcal{D}}\vspace{0.2cm}\\
\hspace{-0.15cm}v_{k}^2 \dfrac{\partial W}{\partial \bfE}\left(\bfE(\bfu_{k})\right)\bfN=\overline{\bfs}(\bfX,t_k),&  \bfX\in\partial\Omega_{\mathcal{N}}
\end{array}\right. \label{BVP-u-theory}
\end{equation}
and
\begin{equation}
\left\{\begin{array}{ll}
\hspace{-0.15cm} {\rm Div}\left[\varepsilon\, \delta^\varepsilon G_c \nabla v_k\right]=\dfrac{8}{3}v_{k} W(\bfE(\bfu_k))+\dfrac{4}{3}c^\varepsilon_\texttt{e}(\bfE(\bfu_k),v_k)-\dfrac{\delta^\varepsilon G_c}{2\varepsilon}+\dfrac{8}{3\,\zeta} \, p(v_{k-1},v_k),& \bfX\in \Omega
\vspace{0.2cm}\\
\hspace{-0.15cm}\nabla v_k\cdot\bfN=0,&  \bfX\in \partial\Omega
\end{array}\right. \label{BVP-v-theory}
\end{equation}
with $p(v_{k-1},v_k)=|v_{k-1}-v_k|-(v_{k-1}-v_k)-|v_k|+v_k$. In these equations, $\textbf{b}(\bfX,t)$ is the body force per unit undeformed volume applied throughout the domain $\Omega$ occupied by the body, $\overline{\bfu}(\bfX,t)$ is the displacement applied at the Dirichlet part $\partial\Omega_{\mathcal{D}}$ of the boundary $\partial\Omega$ of $\Omega$, $\overline{\bfs}(\bfX,t)$ is the traction applied over the complementary Neumann part $\partial\Omega_{\mathcal{N}}=\partial\Omega\setminus\partial\Omega_{\mathcal{D}}$ of the boundary, with outward unit normal $\bfN$, $\nabla v_k(\bfX)=\nabla v(\bfX,t_k)$, $\zeta$ is a penalty parameter\footnote{The penalty function $p(v_{k-1},v_k)$ and penalty parameter $\zeta$ in (\ref{BVP-v-theory}) enforce that the phase field remains in the physically admissible range $0\leq v\leq 1$ and that fracture is irreversible. These requirements can be enforced by means of different strategies \citep{Wick15}, the penalty approach spelled out here being one of them.}  such that\footnote{Typically, it suffices to set $\zeta^{-1}=10^4 \delta^\varepsilon G_c/(2\varepsilon)$.} $\zeta^{-1}\gg \delta^\varepsilon G_c/(2\varepsilon)$, while the $\varepsilon$-dependent function $c^\varepsilon_{\texttt{e}}$ and coefficient $\delta^\varepsilon$ are given by expressions that depend on the specific form of the given material strength surface $\mathcal{F}(\boldsymbol{\sigma})=0$. For the case of a Drucker-Prager strength surface (\ref{DP-1}), they read \cite{KKLP24}:
\begin{equation*}
\left\{\hspace{-0.15cm}\begin{array}{l}
c^\varepsilon_{\texttt{e}}(\bfE(\bfu),v)=\beta^\varepsilon_2 v^2\sqrt{J_2}+
\beta^\varepsilon_1 v^2 I_1-v\left(1-\sgn(I_1)\right)W(\bfE(\bfu))\vspace{0.2cm}\\
\delta^\varepsilon=\left(\dfrac{\sts+(1+2\sqrt{3})\,\shs}{(8+3\sqrt{3})\,\shs}\right)\dfrac{3 G_c}{16W_{\texttt{ts}}\varepsilon}+\dfrac{2}{5}
\end{array}\right. 
\end{equation*}
with $I_1=E/(1-2\nu)\,{\rm tr}\,\bfE(\bfu)$ and $J_2=E^2/(2(1+\nu)^2){\rm tr}\,\bfE^2_D(\bfu)$, where $\bfE_D(\bfu)=\bfE(\bfu)-\frac{1}{3}({\rm tr}\,\bfE(\bfu))\bfI$, $\beta^\varepsilon_1=\delta^\varepsilon G_c/(8\shs\varepsilon)-2W_{\texttt{hs}}/(3\shs)$, $\beta_2=\sqrt{3}(3\shs-\sts)\delta^\varepsilon G_c/(8\shs\sts\varepsilon)+2W_{\texttt{hs}}/(\sqrt{3}\shs)-2\sqrt{3}W_{\texttt{ts}}/\sts$, $W_{\texttt{ts}}=\sts^2/(2E)$, $W_{\texttt{hs}}=3(1-2\nu)\shs^2/(2E)$, and $\shs=2\scs\sts/(3(\scs-\sts))$.

From a physical point of view, we remark that equations (\ref{BVP-u-theory}) are nothing more than the equations of linear elastostatics for the displacement field $\bfu(\bfX,t)$, while equations (\ref{BVP-v-theory}) are the evolution equations for the phase field $v(\bfX,t)$. The latter dictate that the phase field may decrease from its initial state ($v=1$) toward a fractured state ($v=0$) only at material points $\bfX$ where the strength surface (\ref{F-strength}) has been exceeded. Within such regions of strength violation, the decrease in $v$ proceeds according to the minimization of the sum of the elastic and (regularized) surface energies. Put differently,  equations (\ref{BVP-u-theory})-(\ref{BVP-v-theory}) establish that: \emph{an elastic brittle body subjected to mechanical loads undergoes a combined process of elastic deformation and fracture, one where cracks nucleate and propagate strictly within regions where the material's strength surface has been exceeded following an evolution governed by the minimization of the sum of the elastic and surface energies.}

From a mathematical point of view, the equations (\ref{BVP-u-theory})$_1$ and (\ref{BVP-v-theory})$_1$ are second-order elliptic PDEs for $\bfu(\bfX,t)$ and $v(\bfX,t)$, respectively. They are thus amenable to numerical solution via the standard finite element (FE) method. FE implementations are currently available in the open-source platforms FEniCS and MOOSE.\footnote{\url{http://pamies.cee.illinois.edu/repositories/}} Over the past eight years, as alluded to above, these implementations have been deployed to simulate fracture nucleation and propagation in a wide array of hard and soft materials, across diverse specimen geometries and loading conditions. Direct comparisons of these simulations with experiments demonstrate that the phase-field theory (\ref{BVP-u-theory})–(\ref{BVP-v-theory}), and its finite-elasticity counterpart, appear to explain, describe, and predict nucleation and propagation of fracture in nominally elastic brittle materials. In Section \ref{Sec: Demonstration}, using the material constants extracted from the tests proposed in Section \ref{Sec: The experiments}, we generate phase-field predictions and compare them directly with experimental observations of crack nucleation and propagation in unnotched and notched mortar beams under four-point and three-point bending. The FEniCS codes utilized to generate all the simulations presented in this work have been made available on GitHub.\footnote{\url{http://pamies.cee.illinois.edu/repositories/}}

\section{The proposed experiments}\label{Sec: The experiments}

This section details the three tests and their associated analyses proposed to determine the macroscopic material properties that govern crack nucleation and propagation in cementitious materials. The first test, presented in Subsection \ref{Test 1},  is that of the uniaxial compression of a cylindrical specimen. It is utilized to extract the elastic properties, namely, the Young's modulus $E$ and Poisson's ratio $\nu$, as well as the uniaxial compressive strength $\scs$. Subsection \ref{Test 2} presents the second test, a Brazilian fracture test, conducted on a disk using flat platens, which serves to extract the uniaxial tensile strength $\sts$. Knowledge of the uniaxial compressive and uniaxial tensile strengths then allows for the estimation of the material's strength surface $\mathcal{F}(\boldsymbol{\sigma})=0$ via interpolation. In this work, we propose a Drucker-Prager fit (\ref{DP-1}) as the simplest valid interpolation.\footnote{Of course, more sophisticated interpolations are possible, if one is willing to perform the additional testing required to identify further points on the surface $\mathcal{F}(\boldsymbol{\sigma})=0$.} Finally, the third experiment, detailed in Subsection \ref{Test 3}, is the wedge split test on a notched cube, which provides a direct measurement of the fracture toughness $G_c$. 

For demonstration purposes, we carry out the proposed tests and present their analyses for a mortar mixture that has been recently developed for 3D printing \cite{Manaugh2025}. In particular, the mixture consists of Type IL cement with water/cement weight ratio of 0.38 and a 48\% volume fraction of two blended river sands, at a 40/60 blend by weight, with fineness moduli of 3.44 and 2.49. Given the maximum heterogeneity size of 4.75 mm, all tests were conducted on specimens with a minimum dimension of 5 cm. This provided a sufficient separation of length scales to treat the mortar mixture as effectively homogeneous. All the experimental data presented in this work have been made available on the Illinois Data Bank.

\subsection{Uniaxial compression of a cylindrical specimen to extract $E$, $\nu$, and $\sigma_{\texttt{\emph{cs}}}$}\label{Test 1}

\begin{figure}[t!]
\centering
\includegraphics[width=0.85\linewidth]{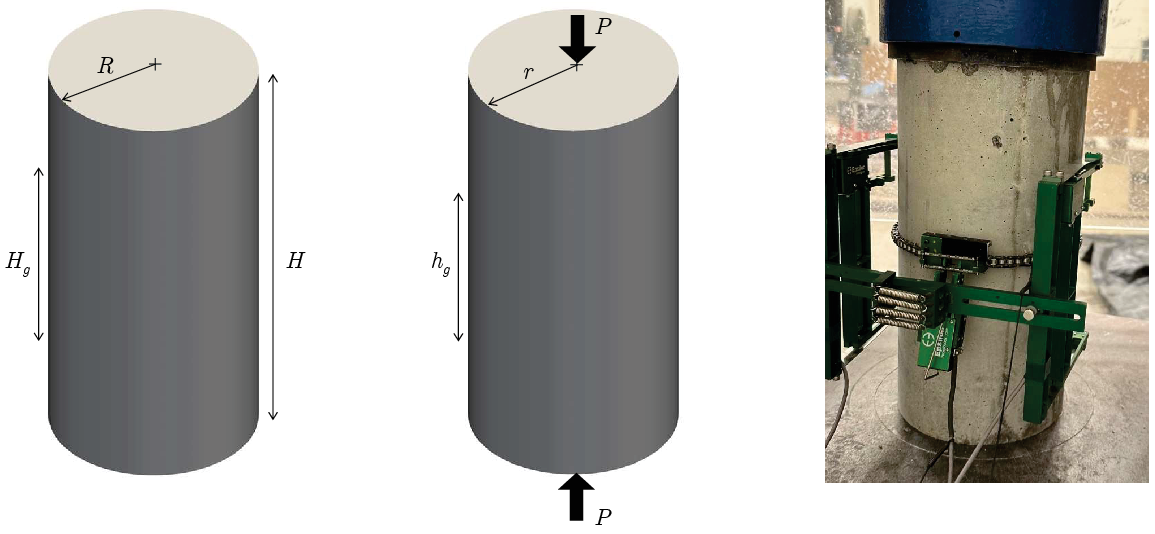}
\caption{{\small Schematics of the geometry of the specimen and the applied loading for the proposed test to extract the Young's modulus $E$, the Poisson's ratio $\nu$, and the uniaxial compressive strength $\scs$. The figure also includes a picture of the setup for the tests performed on mortar ($H=30$ cm, $H_g=15$ cm, and $R=7.5$ cm).}}\label{Fig1}
\end{figure}

The specimen geometry and loading conditions for the first proposed test are illustrated in Fig.~\ref{Fig1}. This test was selected for several practical advantages. 

First, by utilizing a cylinder of height $H$ and circular cross-section radius $R<H$, the geometry avoids sharp corners that are prone to undesirable defects and ensures that the specimen is easy to cast. An additional benefit is that this cylindrical shape is typical of the specimens that are commonly quarried or cored from existing structures for estimating in-situ properties. The geometry is also readily scalable, a critical feature for maintaining the necessary separation of length scales between the specimen dimensions and the largest underlying heterogeneities. For the mortar tested as an example in this work, we used specimens with $H=30$ cm and $R=7.5$ cm so that the smallest specimen dimension, $2R=15$ cm, is more than ten times larger than the maximum aggregate size of 4.75 mm.

The test can be carried out with standard loading frames and flat platens. Furthermore, all necessary data --- including the axial deformation $h_g$ of the gauge section, the applied force $P$, and the radial deformation from the Poisson effect --- can be captured with standard instrumentation.

Finally, the measurements obtained from this test allow for the direct extraction of the target  material properties. Specifically, the Young's modulus is determined from the slope, $E=\Delta S/((H_g-h_g)/H_g)$, of the global stress $S=P/(\pi R^2)$ versus the axial strain $-(h_g-H_g)/H_g$ in the gauge section prior to fracture. The Poisson's ratio $\nu$ is computed from the ratio of radial to axial strain, $\nu=-((r-R)/R)/((h_g-H_g)/H_g)$, also prior to fracture, while the peak global stress $S_{max}$ defines the uniaxial compressive strength $\scs$.

%
%

%
\begin{figure}[t!]
\centering
\includegraphics[width=0.9\linewidth]{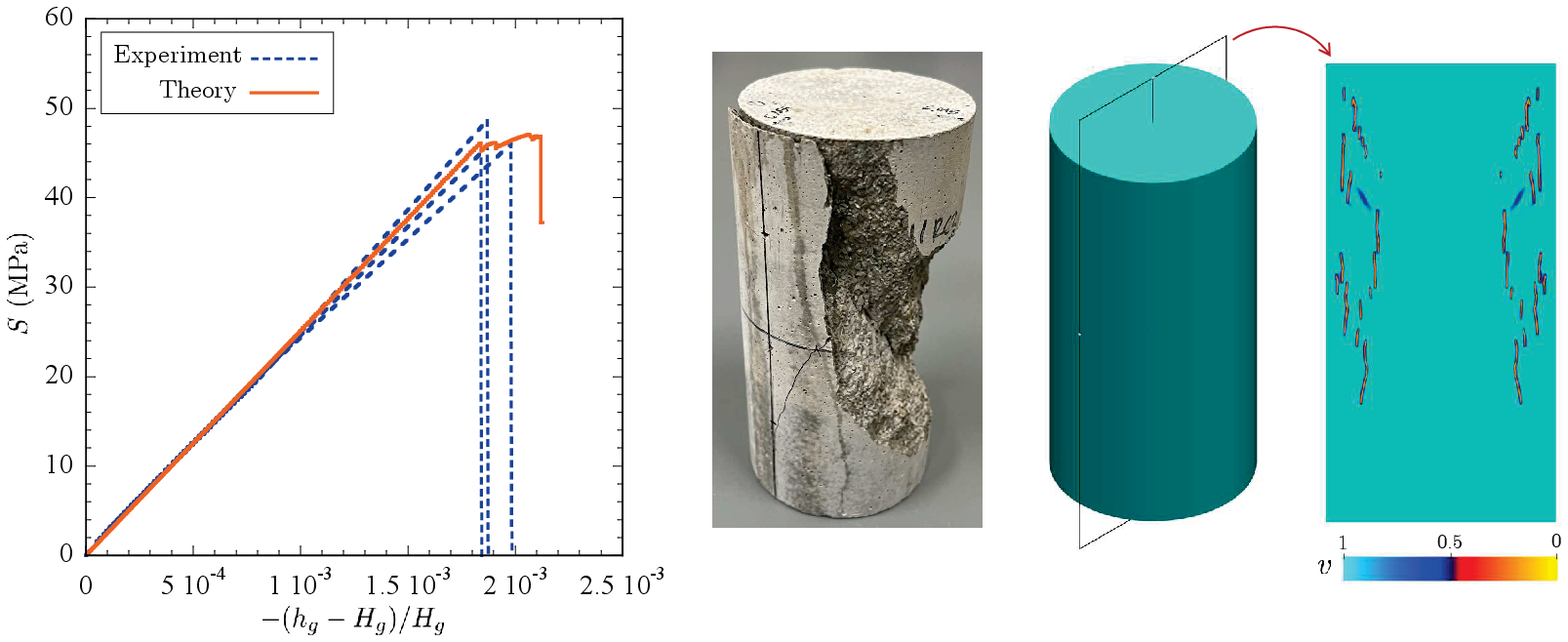}
\caption{{\small Response from three uniaxial compression tests on mortar. The results show the global stress $S=P/(\pi R^2)$ versus the axial strain $-(h_g-H_g)/H_g$ and an image of one of the specimens after crack nucleation at $S=S_{max}$. For direct comparison, the corresponding predictions generated from the phase-field theory (\ref{BVP-u-theory})-(\ref{BVP-v-theory}) are also included.}}\label{Fig2}
\end{figure}
%

%
\begin{table}[H]\centering
\caption{Measurements from the uniaxial compression tests on mortar and the corresponding extracted values for $E$, $\nu$, and $\scs$.}
\begin{tabular}{r|ccc}
\toprule
                              & $\frac{H_g \Delta S}{(H_g-h_g)}=E$ (GPa)            &            $-\frac{H_g(r-R)}{R(h_g-H_g)}=\nu$        &               $S_{max}=\scs$ (MPa)      \\
\midrule
Test 1                        &  24.4                &      0.16               &                 44.9               \\
\midrule
Test 2                        & 24.5                 &      0.16               &                 50.2             \\
\midrule
Test 3                        & 24.6                 &      0.18               &                 49.1               \\
\midrule
Test Average                  &  24.5                &      0.17               &                 48.1                 \\
\midrule
Test Coefficient of Variation           &     0.41\%             &     6.93\%                &        5.82\%                         \\
\bottomrule
\end{tabular} \label{Table1}
\end{table}

By way of an example, Fig.~\ref{Fig2} presents the results obtained from three uniaxial compression tests carried out on mortar. In particular, the plot shows the global stress  $S=P/(\pi R^2)$ as a function of the axial strain $-(h_g-H_g)/H_g$, while the image shows one of the specimens after crack nucleation at $S=S_{max}$. The material properties extracted from these tests are reported in Table \ref{Table1}, alongside their average value and coefficient of variation. 

For direct comparison, we have included in Fig.~\ref{Fig2} the predictions generated by the phase-field theory (\ref{BVP-u-theory})-(\ref{BVP-v-theory}) based on the materials properties --- rounded off to $E=25$ MPa, $\nu=0.17$, $\sts=5$ MPa, $\scs=48$ MPa, $G_c=20$ N/m and summarized in Table \ref{Table4} below --- extracted from this and the two remaining proposed tests. The comparisons between the theoretical predictions and the experimental results in this and subsequent figures (Figs.~\ref{Fig4}, \ref{Fig6}, and \ref{Fig7}) in this section speak for themselves.

\subsection{The Brazilian test on a disk using flat platens to extract $\sigma_{\texttt{\emph{ts}}}$}\label{Test 2}

%
\begin{figure}[b!]
\centering
\includegraphics[width=0.8\linewidth]{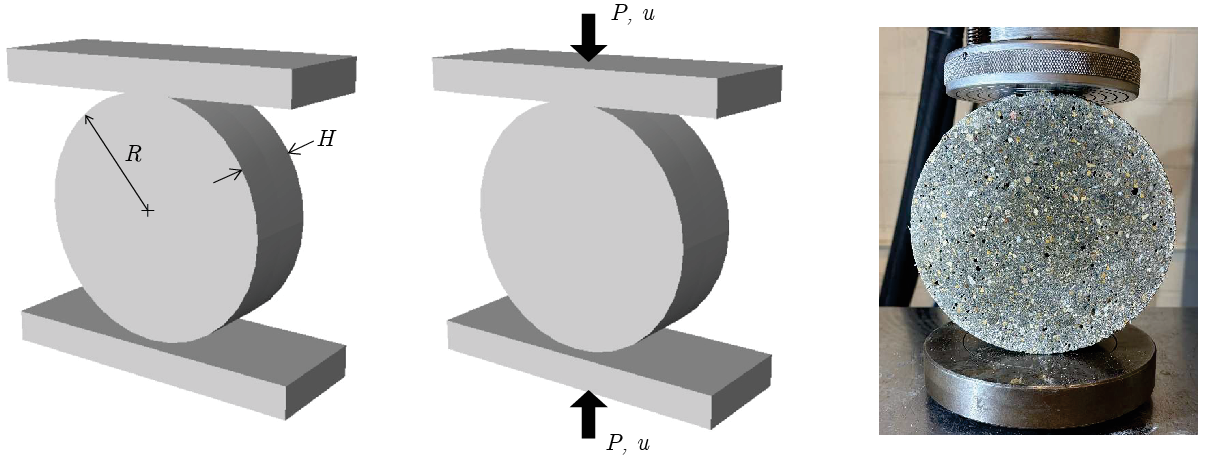}
\caption{{\small Schematics of the geometry of the specimen and the applied loading for the proposed test to extract the uniaxial tensile strength $\sts$. The figure also includes a picture of the setup for the tests performed on mortar ($H=5$ cm and $R=7.5$ cm).}}\label{Fig3}
\end{figure}
%

Figure \ref{Fig3} illustrates the specimen geometry and loading conditions for the second proposed test, a Brazilian test on a disk of height $H$ and radius $R\geq 1.5H$ using flat platens, which was similarly selected for its practical advantages.

Introduced by Lobo Carneiro \cite{Carneiro1943} in the 1940s, the Brazilian test has become a popular test for indirectly probing the tensile strength not only of cementitious materials, but also rocks, ceramics, and other nominally elastic brittle materials with high compressive-to-tensile strength ratios. Its popularity is largely due to its simplicity: it requires minimal specimen preparation (a cylindrical specimen of circular cross section) and straightforward loading between two platens, eliminating the need for specialized grips or fixtures. However, this ease of use is balanced by the complexity of interpreting the results. Because the test induces a non-uniform triaxial stress state rather than uniform uniaxial tension, the measurements of the applied displacement $u$ and resulting force $P$ require careful analysis to be transcribed into material properties. In a recent contribution, Kumar et al. \cite{KLDLP24} have provided a complete analysis of the test, elucidating the shortcomings of previous investigations that rely on purely elastic (predominantly 2D) stress analyses which incorrectly equate fracture nucleation with the maximum principal stress exceeding a threshold at an individual material point. Moreover, for the case of disks with $R\geq 1.5H$ that are compressed between flat platens, these authors have worked out a formula for the uniaxial tensile strength $\sts$ of the material directly in terms of the geometry of the specimen, the maximum force $P_{max}$ recorded in the test, and the uniaxial compressive strength $\scs$. The formula reads
\begin{equation}\label{sts-Brazilian}
\sigma_{\texttt{ts}}=f\left(P_{max},\scs\right)\dfrac{P_{max}}{\pi R H},\qquad f\left(P_{max},\scs\right)=\dfrac{\left(\sqrt{13}-2\right)\dfrac{\pi R H}{P_{max}}\scs}{\dfrac{2\pi R H}{P_{max}}\scs-\sqrt{13}-2}.
\end{equation}

An additional benefit of this test is that the specimen can be cut from samples already cast or cored for uniaxial compression tests. This, in turn, makes the geometry easily scalable, permitting the dimensions to be adjusted to ensure a proper separation of length scales. For the mortar tested here, we employed specimens with $H=5$ cm and $R=7.5$ cm. Maintaining the height at 5 cm ensures that the smallest specimen dimension is approximately ten times the maximum aggregate size of 4.75 mm.

\begin{figure}[t!]
\centering
\includegraphics[width=0.9\linewidth]{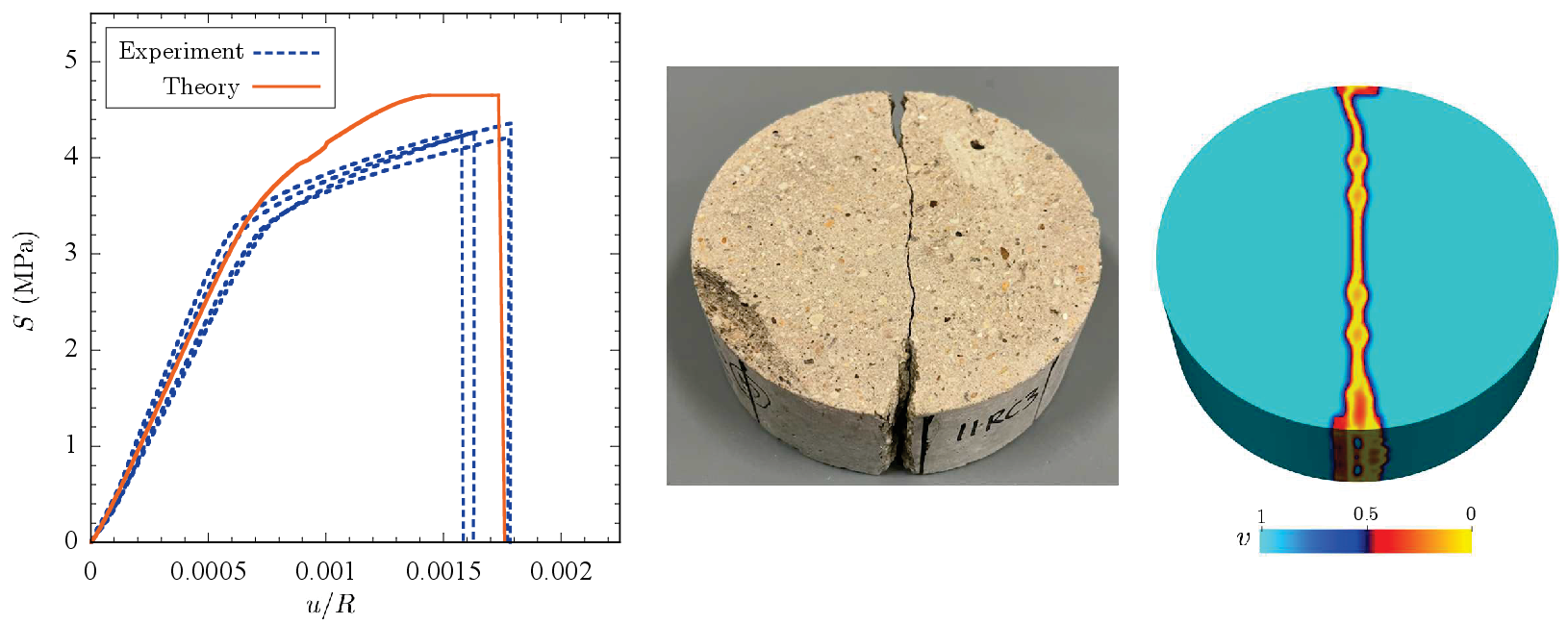}
\caption{{\small Response from four Brazilian tests on mortar. The results show the global stress $S=P/(\pi R H)$ versus the global strain $u/R$ and an image of one of the specimens after crack nucleation at $S=S_{max}$. For direct comparison, the corresponding predictions generated from the phase-field theory (\ref{BVP-u-theory})-(\ref{BVP-v-theory}) are also included.}}\label{Fig4}
\end{figure}

The results of four Brazilian tests performed on mortar are shown in Fig.~\ref{Fig4}. Specifically, the global stress, defined as $S=P/(\pi R H)$, is plotted as a function of the global strain, $u/R$. The accompanying image illustrates a representative specimen after crack nucleation, which occurs at $S=S_{max}$. The resulting material properties are summarized in Table \ref{Table2}, which includes individual test results, their mean values, and the associated coefficients of variation. For direct comparison, the predictions generated by the phase-field theory (\ref{BVP-u-theory})-(\ref{BVP-v-theory}) are also included in Fig.~\ref{Fig4}.

%
%

%
\begin{table}[t!]\centering
\caption{Measurements from the Brazilian tests on mortar and the corresponding values of $\sts$ extracted via formula (\ref{sts-Brazilian}).}
\begin{tabular}{r|cc}
\toprule
                              & $\dfrac{P_{max}}{\pi R H}$ (MPa)        & $\sts$ (MPa)    \\
\midrule
Test 1                        &  4.58          &                5.02                  \\
\midrule
Test 2                        &  4.45          &                4.82                   \\
\midrule
Test 3                        &  4.46          &                4.83                    \\
\midrule
Test 4                        &  4.45          &                4.82                    \\
\midrule
Test Average                  &  4.48          &                4.87                     \\
\midrule
Test Coefficient of Variation    &  1.42\%    &                 2.02\%                    \\
\bottomrule
\end{tabular} \label{Table2}
\end{table}

\subsection{The wedge split test on a notched cube  to extract $G_c$}\label{Test 3}

\begin{figure}[t!]
\centering
\includegraphics[width=0.9\linewidth]{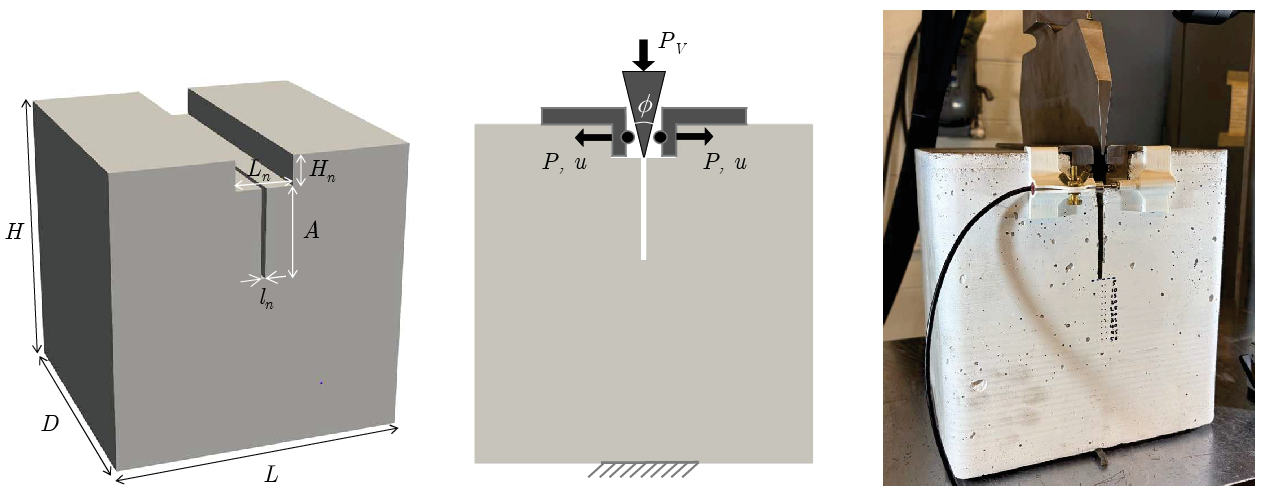}
\caption{{\small Schematics of the geometry of the specimen and the applied loading ($\phi=10^\circ$) for the proposed test to extract the fracture toughness $G_c$. The figure also includes a picture of the setup for the tests performed on mortar ($L=H=D=25$ cm,  $L_n=4.8$ cm, $H_n=2.5$ cm, $l_n=0.4$ cm, $A=7.5$ cm).}}\label{Fig5}
\end{figure}

Finally, Fig.~\ref{Fig5} illustrates the specimen geometry and loading conditions for the third proposed test, a wedge split test on a notched cube of sides $L=H=D$, insert width $L_n$, insert height $H_n$, notch width $l_n$, and  notch length $A$ loaded by means of a wedge with a small angle $\phi$.

%
%
%
\begin{figure}[t!]
\centering
\includegraphics[width=0.9\linewidth]{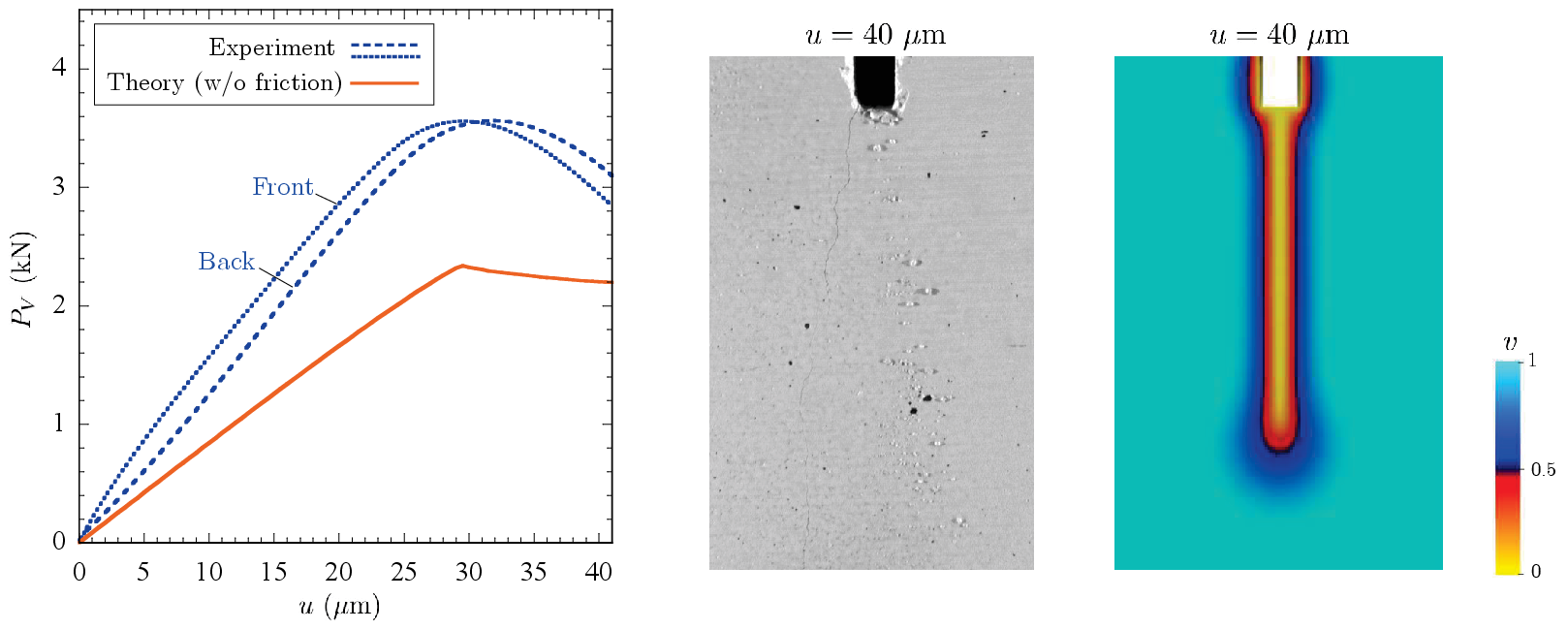}
\caption{{\small Response from one of the wedge split tests on mortar. The results show the global vertical force $P_V$ versus the horizontal displacement $u$, measured at both the front and back faces, and a close-up image of the crack growth on the front face at $u=40$ $\mu$m. For direct comparison, the corresponding predictions generated from the phase-field theory (\ref{BVP-u-theory})-(\ref{BVP-v-theory}) are also included for the idealized case when there is no friction between the wedge and the fixture so that $P=P_V/(2\tan(\phi/2))$.}}\label{Fig6}
\end{figure}
%
%
\begin{figure}[t!]
\centering
\includegraphics[width=0.9\linewidth]{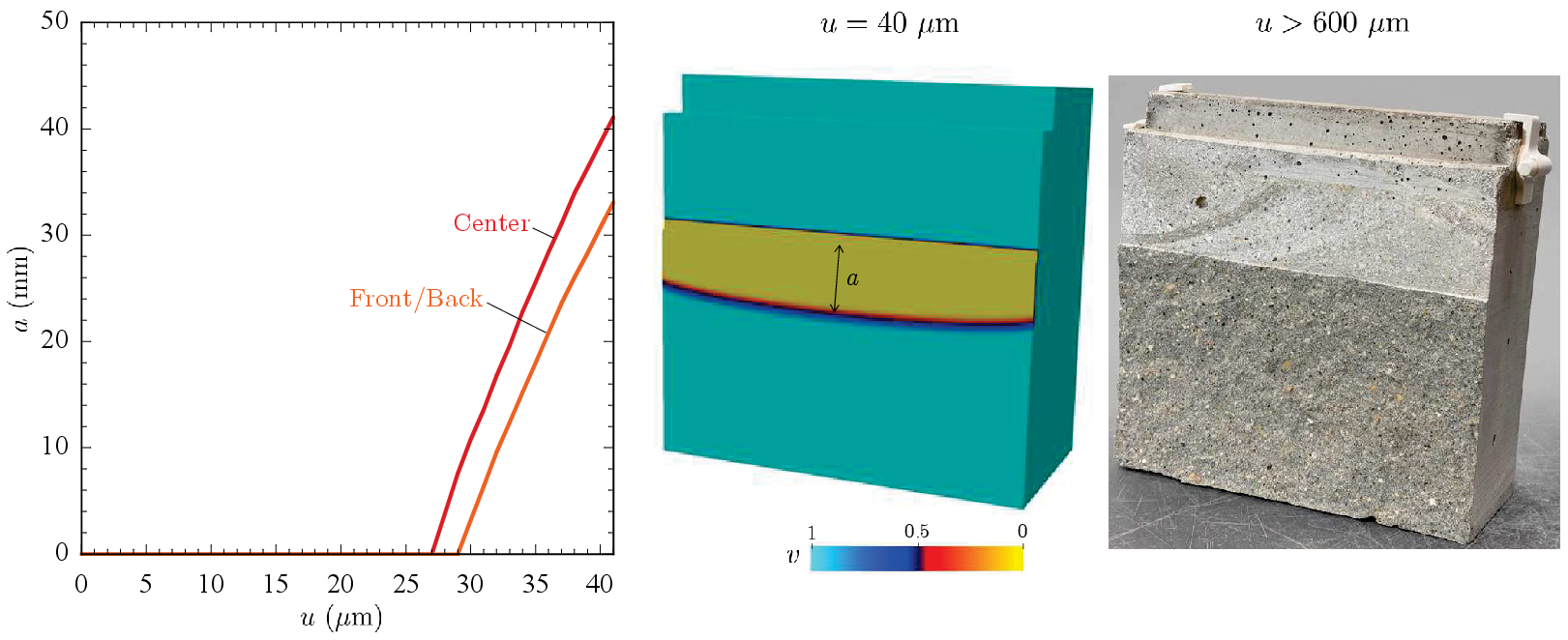}
\caption{{\small Predictions generated from the phase-field theory (\ref{BVP-u-theory})-(\ref{BVP-v-theory}) for the crack growth in the wedge split tests on mortar. The line plot presents the crack length $a$, at the center and at the front and back faces of the specimen, as a function of the horizontal displacement $u$, while the contour plot shows the phase field $v$ at $u=40$ $\mu$m, over the right half of the specimen for better visualization. For comparison, an image of the right half of a fully split specimen (which occurs at around $u=600$ $\mu$m) is also included.}}\label{Fig7}
\end{figure}
%
%
\begin{table}[t!]\centering
\caption{Measurements from the wedge split tests on mortar and the corresponding values of $G_c$ extracted via formula (\ref{Gc-formula}).}
\begin{tabular}{r|cccc}
\toprule
                              & Front $u_c$ ($\mu$m)       &    Back $u_c$ ($\mu$m)    &     Average $u_c$ ($\mu$m)  &     $G_c$ (N/m)      \\
\midrule
Test 1                        &  28.7                        &      31.9                  &                 30.3       &        19.4           \\
\midrule
Test 2                        & 30.0                          &      34.5                    &                 32.3       &     22.1              \\
\midrule
Test 3                        & 31.7                         &      31.9                     &                 31.8       &      21.4              \\
\midrule
Test Average                  &                           &                           &                 31.5      &         21.0           \\
\midrule
Test Coefficient of Variation      &                     &                           &                 3.31\%       &     6.68\%              \\
\bottomrule
\end{tabular} \label{Table3}
\end{table}

The wedge split test --- originally introduced by Linsbauer and Tschegg \cite{Linsbauer1986} and later refined and extended by Br\"uhwiler and Wittmann \cite{Wittmann1990} in the 1980s --- is arguably the most expedient existing test to grow a stable crack in cementitious materials and thereby determine their fracture toughness $G_c$ in a robust and effective manner; hence, its adoption in this work. This is because the specimen is straightforward to cast and load and, more importantly, readily scalable. Again, such scalability is essential for characterizing cementitious materials, whose large aggregates require the use of relatively large specimens to ensure separation of length scales, and is difficult to achieve with competing tests that yield stable crack growth, such as the double cantilever beam test \cite{Gillis1964,Rossi1991}. For the mortar tested in this work, we employed specimens with $L=H=D=25$ cm,  $L_n=4.8$ cm, $H_n=2.5$ cm, $l_n=0.4$ cm, and $A=7.5$ cm. Making use of a notch of length $A=7.5$ cm whose front is $H-(H_n+A)=15$ cm from the bottom boundary of the specimen ensures that both the notch and the dimension ahead of its front are sufficiently large, in this case, more than fifteen times the maximum aggregate size of 4.75 mm.

Previous studies have established two primary methodologies for extracting the fracture toughness $G_c$ from the wedge split test. The first approach equates $G_c$ to the work-of-fracture --- calculated as the area under the horizontal force vs. crack opening displacement ($P$ vs. $2u$) curve --- normalized by the surface area of the created crack. This energy-based calculation can be performed either in its totality \cite{Wittmann1990,Rossi1991,Roesler2008} or incrementally \cite{Rieder2001}. The second approach determines numerically the stress intensity factor $K_{IC}$ as a function of the horizontal force $P$ under plane-strain conditions \cite{Roesler2007} and subsequently relates it to $G_c$ via $K_{IC}=\sqrt{E G_c/(1-\nu^2)}$. Both approaches have been routinely implemented assuming negligible friction, whereby the horizontal force $P$ is taken to be given in terms of the vertical force $P_V$ by the relation
\begin{equation*}
P=\dfrac{P_V}{2\tan\dfrac{\phi}{2}}.
\end{equation*}
However, the presence of friction is unavoidable in practice, in which case $P$ is given in terms of $P_V$ and the coefficient of friction $\mu$ at the wedge/insert interface by the more general relation
\begin{equation*}
P=\dfrac{P_V}{2\tan\dfrac{\phi}{2}}-\dfrac{\mu P_V}{1+\mu\sin\phi-\cos\phi}.
\end{equation*}
A quick calculation suffices to realize that even small values of the coefficient of friction, such as $\mu=0.05$, have a very significant impact on the value of the resulting horizontal force $P$. For example, for the wedge angle of $\phi=10^\circ$ used in our experiments on mortar, $P=5.72 P_V$ without friction, whereas $P=3.62 P_V$ assuming a coefficient of friction of $\mu=0.05$. While one might attempt to measure $\mu$ to correct the force calculation, this approach remains impractical because the precise measurement of such low friction coefficients is non-trivial. Furthermore, different minor misalignments of the wedge and specimen can lead to different effective friction coefficients. 

To address these limitations, we propose an alternative approach to extract $G_c$ from the wedge split test that bypasses the inherently unreliable relationship between the vertical force $P_V$ and the horizontal force $P$. The central idea is to relate $G_c$ to the critical horizontal displacement $u_c$ at which the crack starts to grow on the faces of the specimen. Unlike the force $P$, the displacement $u$ can be directly and robustly measured, for instance, via a linear variable differential transducer (LVDT) mounted on the specimen.\footnote{It is recommended that two LVDTs be used, one on the front face of the specimen and another on the back, to monitor potential misalignments in the loading process.} To this end, following a standard methodology, the energy release rate is first computed via FE analysis for a range of elastic properties (Young's moduli and Poisson's ratios) for the specific specimen geometry, and the results are subsequently fitted to an explicit formula. For the specimen geometry employed for mortar in this work ($L=H=D=25$ cm,  $L_n=4.8$ cm, $H_n=2.5$ cm, $l_n=0.4$ cm, $A=7.5$ cm), we have established the following simple explicit formula\footnote{The functional form of this formula applies broadly to specimens of other dimensions; however, the coefficients $\alpha_1$ through $\alpha_5$ are geometry-dependent and must be re-evaluated accordingly.}
\begin{equation}\label{Gc-formula}
G_c=\left(\dfrac{\alpha_1+\alpha_2 \nu^2}{\alpha_3+\alpha_4 \nu^2+\alpha_5\nu^4}\right)\dfrac{E u^2_{c}}{A},\qquad \left\{\hspace{-0.1cm}\begin{array}{l}
\alpha_1=0.1097\vspace{0.2cm}\\
\alpha_2=-0.0211\vspace{0.2cm}\\
\alpha_3=1.7573\vspace{0.2cm}\\
\alpha_4=-1.4055\vspace{0.2cm}\\
\alpha_5=1.6714
\end{array}\right. .
\end{equation}
With the formula established, the remaining task is to identify $u_c$ from experimental data. Numerical simulations reveal that $u_c$ can be estimated as the horizontal displacement $u$ at which the horizontal force $P$ (and, by extension\footnote{Because $P$ and $P_V$ are related monotonically, their peaks must coincide regardless of the magnitude of friction.}, the vertical force $P_V$) reaches its maximum value. Consequently, by identifying the peak in the $P_V$ vs. $u$ curve, $u_c$ can be readily determined and, in turn, $G_c$ can be extracted.

By way of an example, as for the two preceding tests, Fig.~\ref{Fig6} presents the results of one of the three wedge split tests performed on mortar. Specifically, the plot shows the vertical force $P_V$ as a function of the horizontal displacement $u$, measured at the front and back faces of the specimen via LVDTs to monitor for loading misalignment. The image in Fig.~\ref{Fig6} provides a close-up of the notch front at $u=40$ $\mu$m, showing the presence of a crack --- faintly visible due to its small opening --- that nucleated from the notch front and propagated downwards. The critical displacement values $u_c$ for all three tests --- including their average, coefficient of variation, and the resulting fracture toughness $G_c$ extracted from formula (\ref{Gc-formula}) --- are listed in Table \ref{Table3}. For direct comparison, the corresponding predictions derived from the phase-field theory (\ref{BVP-u-theory})–(\ref{BVP-v-theory}) are again included in Figure \ref{Fig6}. These theoretical predictions correspond to the idealized case wherein there is no friction between the wedge and the inserts that transfer the vertical force $P_V$ into the horizontal splitting force $P$.

Several key observations can be made from Fig.~\ref{Fig6}. First, the initial experimental values of the vertical force $P_V$ prior to fracture are significantly larger than those predicted theoretically under the assumption of zero friction. This discrepancy indicates that friction is present\footnote{By assuming a coefficient of friction of $\mu=0.055$, the theoretical response would match the initial slopes of the experimental $P_V$ vs. $u$ curves.} in the test and that its impact is substantial, as expected. Second, the difference between the horizontal displacements measured at the front and back faces of the specimen is relatively small, indicating that loading misalignment was minimal. Third, the experimental vertical force $P_V$ exhibits a distinct peak at a horizontal displacement of approximately $u_c=30.7$ $\mu$m. This peak signals that a crack has already nucleated and propagated to a size sufficient to induce system softening, thereby resulting in the observed maximum force. While a direct visual of the process of crack nucleation and initial propagation is not possible from the experiment, it is provided by the simulation. As illustrated in Fig.~\ref{Fig7}, the latter shows that the crack nucleates in the center of the specimen and subsequently propagates downwards and laterally. Consistent with the formula (\ref{Gc-formula}), the vertical force reaches its peak value only when the crack reaches the front and back faces of the specimen.

\section{Validation results for different structures and loadings}\label{Sec: Demonstration}

%
\begin{figure}[b!]
\centering
\includegraphics[width=0.95\linewidth]{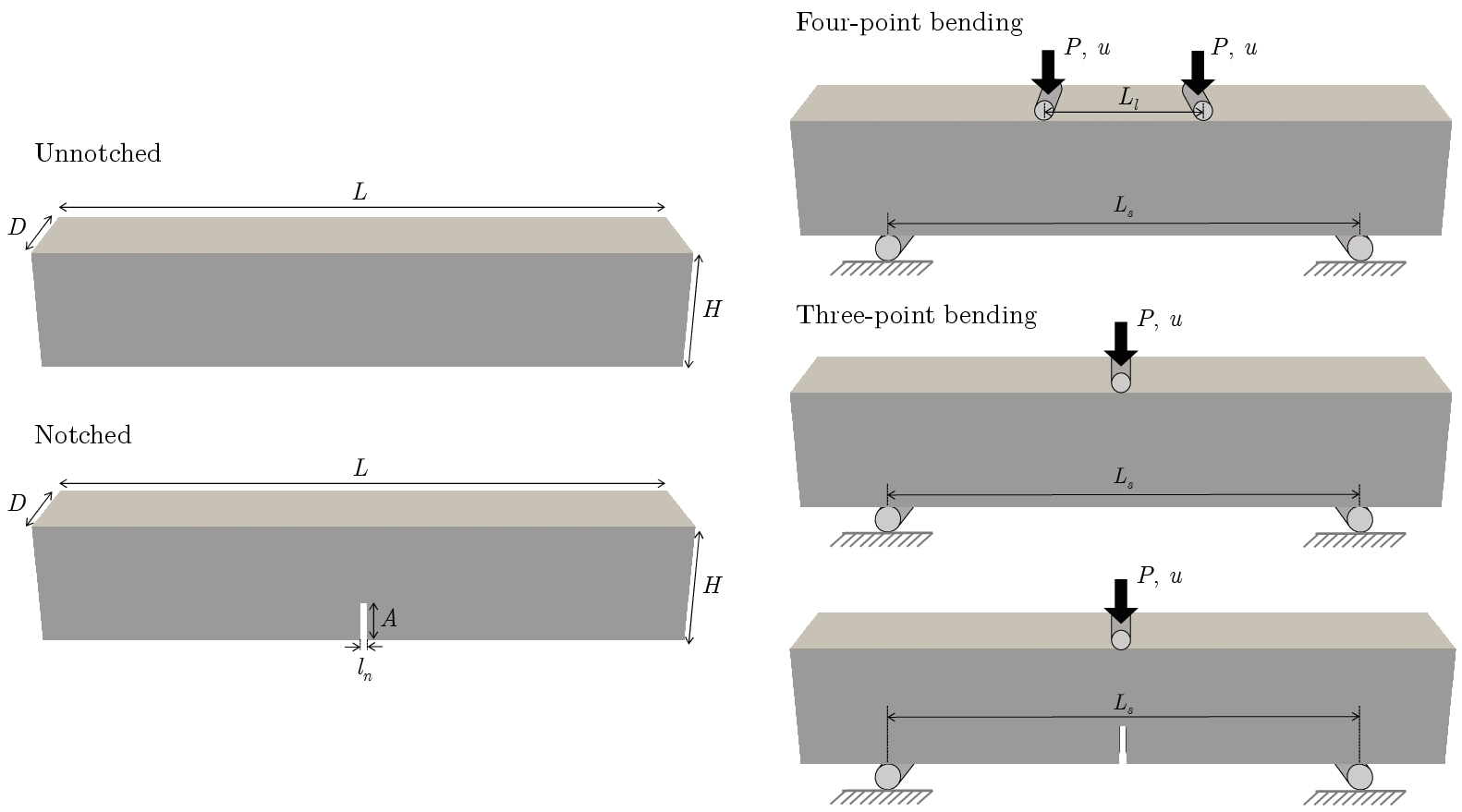}
\caption{{\small Schematics of the geometry of the unnotched ($L=40$ cm, $H=D=7.5$ cm) and notched ($L=40$ cm, $H=D=7.5$ cm, $A=2.5$ cm, $l_n=0.4$ cm) beams and the four-point ($L_s=30$ cm, $L_l=10$ cm) and three-point ($L_s=30$ cm) bending applied to them.}}\label{Fig8}
\end{figure}
%

In this section, we demonstrate that the macroscopic fracture behavior of the mortar studied in this work --- as a representative cementitious material --- is fully characterized by the material constants summarized in Table~\ref{Table4}. Derived from the three tests proposed in the preceding section, these constants govern crack nucleation and propagation in any structure made of this mortar under arbitrary monotonic quasi-static loading (provided that separation of length scales is maintained).
\begin{table}[H]\centering
\caption{Material constants for the mortar mixture studied in this work, extracted from the three proposed tests.}
\begin{tabular}{ccccc}
\toprule
 $E$ (GPa)                 &    $\nu$    &    $\sts$ (MPa)    &     $\scs$ (MPa)  &     $G_c$ (N/m)      \\
\midrule
25                       &       0.17    &      5              &         48       &        20           \\
\bottomrule
\end{tabular} \label{Table4}
\end{table}

Recall that the material constants in Table \ref{Table4} fully describe the elastic energy density (\ref{W-elastic}), the strength surface (\ref{F-strength}) --- here assumed to be of the Drucker-Prager form (\ref{DP-1}) --- and the fracture toughness (\ref{Gc-toughness}) of the material. The phase-field theory (\ref{BVP-u-theory})--(\ref{BVP-v-theory}), utilizing these three material inputs, provides a complete framework to characterize and predict fracture nucleation and propagation in any boundary-value problem of interest. We have already presented comparisons in Figs.~\ref{Fig2}, \ref{Fig4}, \ref{Fig6}, and \ref{Fig7} between the theoretical predictions based on the material constants in Table~\ref{Table4} and the three proposed tests themselves. Below, we present additional validation results for three other experiments: unnotched beams under four-point bending (Subsection~\ref{Sec: 4PBT}), unnotched beams under three-point bending (Subsection \ref{Sec: 3PBT}), and notched beams under three-point bending (Subsection \ref{Sec: 3PBT Notched}). Figure \ref{Fig8} illustrates the specific geometry of the specimens, designed to satisfy the required separation of length scales, along with the loading conditions.

The rationale for choosing these bending tests is severalfold. First, they are classical methods for probing tensile strength, dating back to the 19th century \cite{Rankine1858}. Like the Brazilian test, bending configurations are ubiquitous due to their ease of casting and loading. However, this experimental simplicity is countered by the complexity of interpreting the resulting data, as these tests induce a non-uniform triaxial stress state rather than simple, uniform uniaxial tension. Two historically challenging features are particularly noteworthy: ($i$) the observation that four-point bending typically yields a lower modulus of rupture ($S_{max}$) than three-point bending, and ($ii$) the size effect, where larger beams exhibit a smaller modulus of rupture than smaller counterparts with identical span-to-height ratios \cite{Wright1952,Lindner1955,Walker1957}. Following our recent 3D quantitative analysis and explanation of these phenomena \cite{SRLP26}, which show that \emph{all} material constants ($E$, $\nu$, $\sts$, $\scs$, $G_c$) are involved to a significant extent in the fracture of beams under bending, the focus here is to demonstrate that the phase-field theory, using the constants in Table~\ref{Table4}, accurately predicts crack nucleation and propagation across these diverse configurations.

\subsection{Four-point bending tests on unnotched beams}\label{Sec: 4PBT}

%
\begin{figure}[H]
\centering
\includegraphics[width=0.9\linewidth]{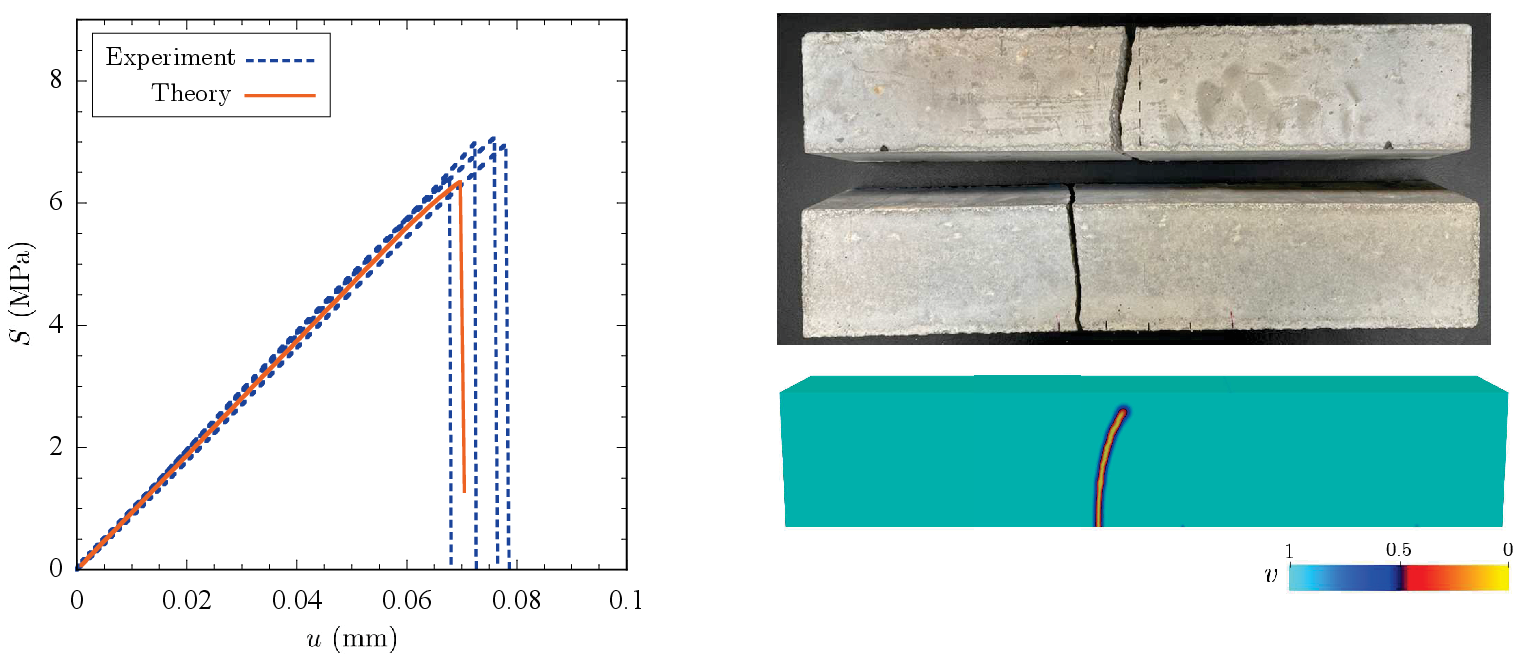}
\caption{{\small Comparison between the theoretical phase-field predictions --- based on the material properties ($E$, $\nu$, $\scs$, $\sts$, $G_c$) extracted from the tests proposed in Section \ref{Sec: The experiments} --- and the experimental results for four-point bending tests on unnotched mortar beams. The line plot presents the global stress $S=3P(L_s-L_l)/(D H^2)$ versus the applied displacement $u$, while the images show select beams immediately before (simulation) and after complete failure (experiments).}}\label{Fig9}
\end{figure}
%

Figure \ref{Fig9} compares the theoretical predictions based on the material constants in Table \ref{Table4} with experimental results from four separate four-point bending tests. The plots show the global stress, $S=3P(L_s-L_l)/(D H^2)$, as a function of the applied displacement $u$. The figure includes images of two specimens after complete failure to illustrate the observed crack paths, alongside the theoretical contour plot predicted immediately prior to failure for direct comparison. 

The main observation is that the theoretical predictions are in good qualitative and quantitative agreement with the experiments regarding both when and where cracks nucleate and propagate. We remark that the simulation did not account for the inherent stochasticity in the uniaxial tensile and compressive strengths, $\sts$ and $\scs$. As shown in \cite{SRLP26}, even a mild stochasticity in these constants leads to variability in the modulus of rupture and the crack nucleation site similar to that observed in the experiments. Table \ref{Table5} lists these experimental results, their average, and the coefficient of variation, alongside the corresponding theoretical predictions. 

Finally, it is worth noting that the exact location of crack nucleation under four-point bending is highly stochastic within the inner loading span, and the resulting crack trajectories can be curved. This contrasts with the fracture behavior under three-point bending, where, as shown next, cracks predominantly nucleate at around the midspan and propagate along a straight, central path.

\begin{table}[t!]\centering
\caption{Critical global stresses and crack location at crack nucleation for the four-point bending tests on unnotched beams.}
\begin{tabular}{r|cc}
\toprule
                              & $S_{max}$ (MPa)      &   \emph{ Crack Distance to Midspan} (cm)       \\
\midrule
Test 1                        &  6.43                &                $+0.3$                                  \\
\midrule
Test 2                        & 6.99                 &                $-4.6$                                 \\
\midrule
Test 3                        & 7.08                 &                $+4.2$                                  \\
\midrule
Test 4                        &  6.96                &                $+2.1$                                  \\
\midrule
Test Average                  &  6.86                &                                                     \\
\midrule
Test Coefficient of Variation      &  4.29\%              &                                         \\
\midrule
\midrule
Theoretical Prediction       &  6.35               &                                                      \\
\bottomrule
\end{tabular} \label{Table5}
\end{table}

\subsection{Three-point bending tests on unnotched beams}\label{Sec: 3PBT}

%
\begin{figure}[H]
\centering
\includegraphics[width=0.9\linewidth]{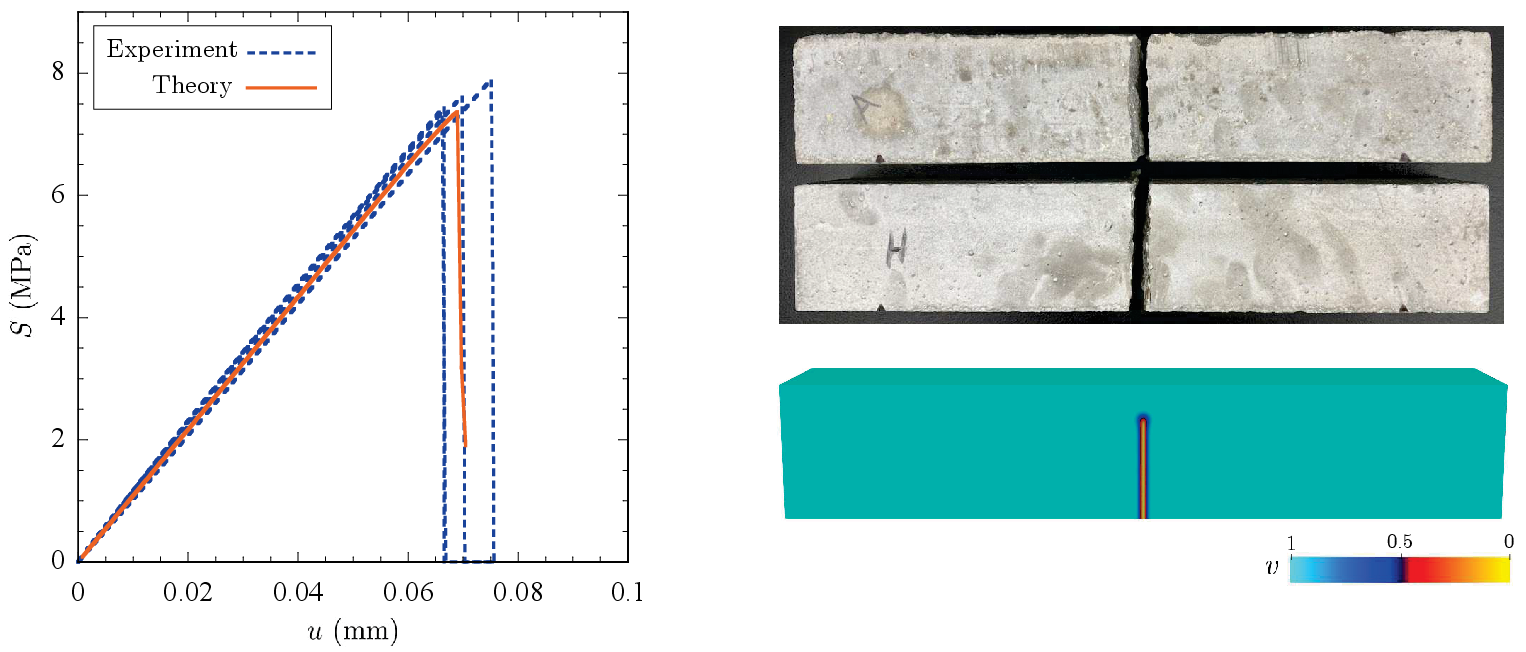}
\caption{{\small Comparison between the theoretical phase-field predictions --- based on the material properties ($E$, $\nu$, $\scs$, $\sts$, $G_c$) extracted from the tests proposed in Section \ref{Sec: The experiments} --- and the experimental results for three-point bending tests on unnotched mortar beams. The line plot presents the global stress $S=3P L_s/(2 D H^2)$ versus the applied displacement $u$, while the images show select beams immediately before (simulation) and after complete failure (experiment).}}\label{Fig10}
\end{figure}
%

Figure \ref{Fig10} presents the comparison between the theoretical predictions, computed using the constants in Table \ref{Table4}, and the experimental data obtained from four separate three-point bending tests on unnotched specimens. The global stress, $S=3P L_s/(2 D H^2)$, is plotted as a function of the applied displacement $u$. Images of two specimens after complete failure are included to show the resulting crack paths, which are presented alongside the theoretical contour plot predicted by the theory immediately before failure for direct comparison.

Consistent with the preceding results, the main observation is that the theoretical predictions are in qualitative and quantitative agreement with the experiments regarding both the onset and trajectory of crack nucleation and propagation. We remark again that the current deterministic simulations do not account for the inherent stochasticity in the uniaxial tensile and compressive strengths. We did so in \cite{SRLP26}, where we showed that incorporating a mild stochasticity into these material constants leads to variability in the modulus of rupture ($S_{max}$) similar to that observed experimentally. Table \ref{Table6} summarizes these experimental results, including their average and coefficient of variation, alongside the corresponding theoretical prediction.

As a final comment, it is interesting to observe that the average modulus of rupture measured from the three-point bending tests ($S_{max}=7.59$ MPa) is approximately 11\% higher than that obtained from the four-point bending tests ($S_{max}=6.86$ MPa). Consistent with this experimental observation, the theory predicts (without strength stochasticity) a comparable difference of 14\%.

\begin{table}[t!]\centering
\caption{Critical global stresses and crack location at crack nucleation for the three-point bending tests on unnotched beams.}
\begin{tabular}{r|cc}
\toprule
                              & $S_{max}$ (MPa)      &   \emph{Crack Distance to Midspan} (cm)       \\
\midrule
Test 1                        &  7.90                &     $-2.3$                                             \\
\midrule
Test 2                        & 7.63                 &     $-1.9$                                             \\
\midrule
Test 3                        & 7.45                 &     $+1.9$                                              \\
\midrule
Test 4                        &  7.37                &     $-0.4$                                             \\
\midrule
Test Average                  &  7.59                &                                                    \\
\midrule
Test Coefficient of Variation      &  3.10\%                &                                                    \\
\midrule
\midrule
Theoretical Prediction       &  7.37                 &                                                      \\
\bottomrule
\end{tabular} \label{Table6}
\end{table}

\subsection{Three-point bending tests on notched beams}\label{Sec: 3PBT Notched}

%
\begin{figure}[H]
\centering
\includegraphics[width=0.9\linewidth]{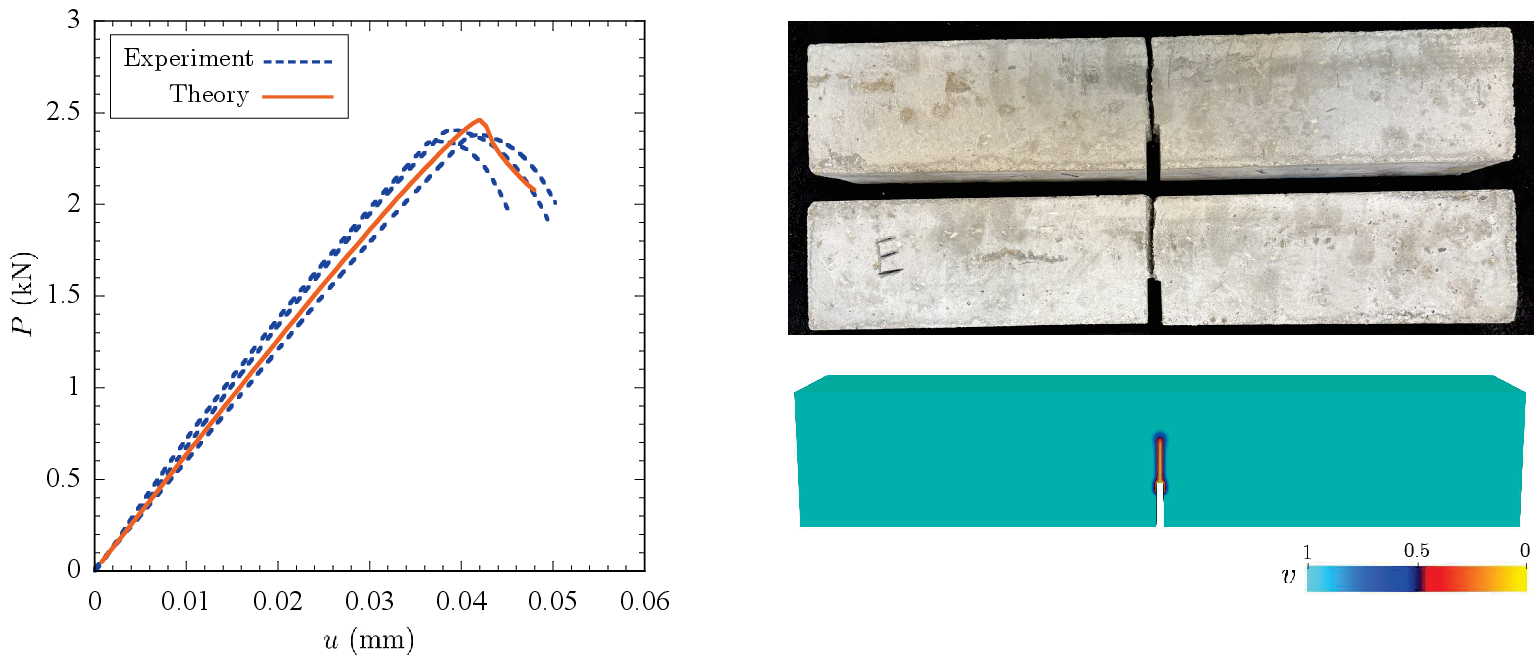}
\caption{{\small Comparison between the theoretical phase-field predictions --- based on the material properties ($E$, $\nu$, $\scs$, $\sts$, $G_c$) extracted from the tests proposed in Section \ref{Sec: The experiments} --- and the experimental results for three-point bending tests on notched mortar beams. The line plot presents the force $P$ versus the applied displacement $u$, while the images show select beams immediately before (simulation) and after complete failure (experiment).}}\label{Fig11}
\end{figure}
%

Lastly, Fig.~\ref{Fig11} and Table \ref{Table7} compare the theoretical predictions, computed using the material constants in Table \ref{Table4}, with the experimental data obtained from three distinct three-point bending tests on notched specimens. In Fig.~\ref{Fig11}, the global force $P$ is plotted as a function of the applied displacement $u$. To illustrate the crack paths, images of two specimens after complete failure are included in the figure. These are displayed alongside the theoretical contour plot predicted by the theory immediately before failure for direct comparison. Additionally, Table \ref{Table7} lists the peak forces $P_{max}$ at which crack nucleation occurred from the notch front.

The theoretical predictions again demonstrate agreement with the experiments, accurately capturing both the nucleation and the subsequent propagation of the cracks.

\begin{table}[t!]\centering
\caption{Critical loads at crack nucleation for the three-point bending tests on notched beams.}
\begin{tabular}{r|c}
\toprule
                              & $P_{max}$ (kN)       \\
\midrule
Test 1                        &  2.39                \\
\midrule
Test 2                        & 2.34                 \\
\midrule
Test 3                        & 2.37                \\
\midrule
Test Average                  &  2.37                \\
\midrule
Test Coefficient of Variation      &  1.06\%             \\
\midrule
\midrule
Theoretical Prediction                        &  2.46              \\
\bottomrule
\end{tabular} \label{Table7}
\end{table}

\section{Final comments}

By leveraging recent advances in fracture mechanics, this work has identified the three fundamental macroscopic properties that govern fracture nucleation and propagation in mortar and concrete. In doing so, it resolves a long-standing ambiguity in the cementitious materials community, where conventional testing has often conflated intrinsic material behavior with extrinsic structural effects.

These properties can be obtained through a minimal and practical experimental framework consisting of three tests: ($i$) uniaxial compression of a cylindrical specimen, ($ii$) the Brazilian test of a disk using flat platens, and ($iii$) the wedge split test on a notched cube. Together, they enable the determination of elastic constants ($E$ and $\nu$), strength parameters ($\sts$ and $\scs$) --- whose interpolation serves to estimate the material strength surface ($\mathcal{F}(\boldsymbol{\sigma})=0$) --- and fracture toughness ($G_c$) using direct force-displacement measurements and simple explicit formulas. Notably, the formula introduced herein for the wedge split test is a novel contribution of this work; unlike previously put forth expressions, it accounts for the unavoidable frictional forces inherent in the wedge-loading mechanism, which otherwise leads to significant overestimations of the fracture toughness.

A key advantage of the proposed framework is its scalability: specimen dimensions can be chosen large enough to sufficiently exceed the characteristic heterogeneity of the material, ensuring that the extracted properties reflect intrinsic behavior. This addresses a common limitation in concrete testing, where undersized specimens often obscure the true material response.

Application to a 3D-printable mortar demonstrates that the measured elasticity, strength, and toughness are sufficient to accurately predict crack nucleation and propagation in both notched and unnotched beams under bending. More broadly, these results indicate that the elasticity, strength, and toughness properties extracted from the proposed suite are sufficient to predict fracture behavior in any structure --- provided a separation of length scales exists --- made of cementitious materials under any monotonic quasi-static loading condition.

\section*{Acknowledgements}

\noindent This work was supported by the National Science Foundation through the Grant DMS--2308169. This support is gratefully acknowledged. The authors also thank Dr. Jordan Ouellet of Newmark Structural Engineering Laboratory at the University of Illinois Urbana-Champaign for his invaluable expertise and assistance with the experiments reported herein.


\bibliographystyle{unsrtnat}
\bibliography{References}

\end{document}